\documentclass[lettersize,journal]{IEEEtran}
\usepackage{amsmath,amsfonts}
\usepackage{algorithm}
\usepackage{algorithmic}
\usepackage{dsfont}
\usepackage{array}
\usepackage[caption=false,font=normalsize,labelfont=sf,textfont=sf]{subfig}
\usepackage{textcomp}
\usepackage{stfloats}
\usepackage{url}
\usepackage{verbatim}
\usepackage{graphicx}
\usepackage{cite}
\usepackage{url}
\usepackage{amsthm}
\usepackage{xcolor}
\usepackage{mathtools}
\usepackage{multirow}

\newcommand{\blue}[1]{\textcolor{black}{#1}}

\newcommand{\dint}[2]{[\![#1 ,#2]\!]}

\newcommand{\Z}{\mathbb{Z}}
\newcommand{\R}{\mathbb{R}}
\newcommand{\Ltn}[1]{\left\lVert #1 \right\rVert _2}
\newcommand{\M}{\mathcal{M}}
\newcommand{\rvect}{\boldsymbol{r}}
\newtheorem{proposition}{Proposition}

\begin{document}

\title{Fully Reversing the Shoebox Image Source Method: From Impulse Responses to Room Parameters}

\author{Tom Sprunck, Antoine Deleforge, Yannick Privat and  Cédric Foy
\thanks{This work was made with the support of the French National Research Agency through project DENISE (ANR-20-CE48-0013).}
\thanks{Antoine Deleforge and Tom Sprunck are with IRMA, CNRS, Universit\'e de Strasbourg, Inria, 67000 Strasbourg, France.}
\thanks{Yannick Privat is with IECL, Universit\'e de Lorraine, CNRS, Inria, BP 70239 54506 Vand\oe uvre-l\`es-Nancy Cedex, France (\texttt{yannick.privat@univ-lorraine.fr})}
\thanks{Yannick Privat is with Institut Universitaire de France (IUF).}
\thanks{Cédric Foy is with UMRAE, Cerema, Univ. Gustave Eiffel, Ifsttar, Strasbourg, 67035, France (cedric.foy@cerema.fr).}
\vspace{-4mm}
}

\markboth{Preprint, October 2024}%
{Shell \MakeLowercase{\textit{et al.}}: A Sample Article Using IEEEtran.cls for IEEE Journals}


\maketitle

\begin{abstract}
We present an algorithm that fully reverses the shoebox image source method (ISM), a popular and widely used room impulse response (RIR) simulator for cuboid rooms introduced by Allen and Berkley in 1979. More precisely, given a discrete multichannel RIR generated by the shoebox ISM for a microphone array of known geometry, the algorithm reliably recovers the 18 input parameters. These are the 3D source position, the 3 dimensions of the room, the 6-degrees-of-freedom room translation and orientation, and an absorption coefficient for each of the 6 room boundaries. The approach builds on a recently proposed gridless image source localization technique combined with new procedures for room axes recovery and first-order-reflection identification. Extensive simulated experiments reveal that near-exact recovery of all parameters is achieved for a 32-element, 8.4-cm-wide spherical microphone array and a sampling rate of 16~kHz using fully randomized input parameters within rooms of size 2$\times$2$\times$2 to 10$\times$10$\times$5 meters. Estimation errors decay towards zero when increasing the array size and sampling rate. The method is also shown to strongly outperform a known baseline, and its ability to extrapolate RIRs at new positions is demonstrated. Crucially, the approach is strictly limited to low-passed discrete RIRs simulated using the vanilla shoebox ISM. Nonetheless, it represents to our knowledge the first algorithmic demonstration that this difficult inverse problem is in-principle fully solvable over a wide range of configurations.
\end{abstract}


\section{Introduction}
\IEEEPARstart{H}{\textit{earing}} \textit{the shape of a room}, or more formally the problem of recovering the properties of a room boundary from the acoustic measurements of one or several sound sources inside of it, is a difficult inverse problem that has intrigued researchers in audio signal processing and room acoustics for many years. Beyond its folklore nature, solutions to this problem could benefit applications in augmented reality \cite{remaggi2019perceived,neidhardt2022perceptual}, room compensation \cite{canclini2012room}, sound field reconstruction \cite{bastine2023room,sundstrom2024optimal}, robotic navigation \cite{krekovic2016echoslam,saqib2020model}, or room acoustic diagnosis \cite{dilungana2022geometry}.

A steady number of approaches have been proposed to tackle different facets of this question along the past two decades \cite{kuster2004acoustic,nava2009situ,antonacci2010geometric,tervo2010estimation,filos2010two,canclini2011direction,dokmanic2011can,nastasia2011localization,canclini2011exact,ribeiro2011geometrically,antonacci2012inference,sun2012localization,filos2012localization,dokmanic2013acoustic,mabande2013room,kowalczyk2013blind,torres2013room,moore2013room,markovic2013estimation,markovic2013soundfield,zamaninezhad2014localization,remaggi20153d,tervo2015direction,antonello2015evaluation,bertin2016joint,crocco2016estimation,jager2016room,rajapaksha2016geometrical,remaggi2016acoustic,el20173d,remaggi2018acoustic,lovedee2019three,di2020blaster,yu2020room,tuna20203d,okawa2021estimation,shlomo2021blind,shlomo2021blindloc,dilungana2022geometry,sprunck2022gridless,tuna2023data,yeon2023rgi,bicer2024data}, but their direct comparison is nearly impossible due to the many variations of the problem that have been considered. Most approaches make use of room impulse responses (RIRs), but some tackle the question \textit{blindly}, namely, with no knowledge of the source signals \cite{tervo2010estimation, filos2010two, canclini2011direction, kowalczyk2013blind, sun2012localization, bertin2016joint, di2020blaster, shlomo2021blind, shlomo2021blindloc}, while other approaches assume that the times of arrival of acoustic reflections are directly available \cite{dokmanic2011can, nastasia2011localization, dokmanic2013acoustic,moore2013room,markovic2013estimation,jager2016room,rajapaksha2016geometrical,dilungana2022geometry}. Most approaches make use of multiple microphones at arbitrary known locations, but some use a single microphone \cite{antonacci2010geometric,dokmanic2011can,moore2013room,markovic2013estimation,zamaninezhad2014localization,tuna20203d,yu2020room,tuna2023data} or specific microphone array geometries such as spherical \cite{sun2012localization,mabande2013room,tervo2015direction,lovedee2019three,shlomo2021blind,shlomo2021blindloc}, circular \cite{canclini2011direction,torres2013room,bicer2024data}, linear \cite{kuster2004acoustic,markovic2013soundfield}, or others \cite{filos2012localization,remaggi20153d,remaggi2018acoustic}. Most approaches use a single sound source, but some use multiple sources \cite{antonacci2010geometric,tervo2010estimation,nastasia2011localization,filos2012localization,remaggi20153d,jager2016room,bertin2016joint,remaggi2016acoustic,remaggi2018acoustic,lovedee2019three,dilungana2022geometry} or a linear loudspeaker array \cite{el20173d,tuna20203d,tuna2023data}. 
Some approaches require specific geometrical assumptions on the setup, such as a rectangular or cuboid ("shoebox") room \cite{filos2012localization,moore2013room,bertin2016joint,yu2020room,sprunck2022gridless,tuna2023data} or having all sources and microphones lying on a plane parallel to both the floor and ceiling \cite{antonacci2010geometric,filos2012localization,antonacci2012inference,tuna20203d,tuna2023data,yeon2023rgi,bicer2024data}. Some approaches focus on the 2D case \cite{filos2010two,canclini2011direction,canclini2011exact,dokmanic2011can,markovic2013estimation,markovic2013soundfield,moore2013room,bertin2016joint,okawa2021estimation} or the 1D case \cite{zamaninezhad2014localization}.
The question of what is to be recovered also varies greatly.
The distance and orientation of one acoustic reflector \cite{tervo2010estimation,canclini2011direction,canclini2011exact,remaggi20153d}, several reflectors \cite{antonacci2010geometric,nastasia2011localization,dokmanic2013acoustic,mabande2013room,jager2016room,remaggi2016acoustic,remaggi2018acoustic,bicer2024data} or all of the reflectors \cite{filos2010two,dokmanic2011can,ribeiro2011geometrically,filos2012localization,antonacci2012inference,moore2013room,zamaninezhad2014localization,rajapaksha2016geometrical,el20173d,lovedee2019three,tuna20203d,tuna2023data,yeon2023rgi} in the room? The unlabelled time differences of arrival \cite{kowalczyk2013blind,crocco2016estimation,di2020blaster}, directions of arrival \cite{sun2012localization, torres2013room,markovic2013soundfield,tervo2015direction}, or 3D positions \cite{shlomo2021blindloc,sprunck2022gridless} of image sources? 
Additional properties of the reflectors such as their absorption \cite{nava2009situ,markovic2013estimation,antonello2015evaluation,bertin2016joint,yu2020room,shlomo2021blind, okawa2021estimation, dilungana2022geometry,sprunck2022gridless} or their size \cite{remaggi2018acoustic}? Finally, the considered noise, sensor, and sound propagation models may differ widely across existing approaches. 

A number of successes have been obtained over the years, including demonstrations on real measured acoustic data \cite{ribeiro2011geometrically,sun2012localization,dokmanic2013acoustic,torres2013room,remaggi2016acoustic,remaggi2018acoustic,lovedee2019three,yu2020room,tuna2023data}. However, because there seems to be nearly as many ways of framing the question as there are research articles on the topic, the more fundamental question of \textbf{whether the problem is solvable at all for a clearly specified and broad enough set of assumptions} remains largely open to date.
In this article, we do not introduce a solution that is readily applicable to real data, but turn our attention towards this more fundamental question instead. To this aim, we focus on a simple but fully specified forward room acoustic model, namely, the well-known shoebox image source method (ISM) proposed by Allen and Berkley in 1979 \cite{allen1979image}. We then frame the question as one of \textit{algorithmic reversibility}, namely, is there an algorithm that, given the output of the shoebox ISM, can reliable recover all of its input? More precisely, given a discrete, ideally low-passed, multichannel RIR generated by the ISM for a microphone array of known geometry, we ask whether the following 18 parameters can be recovered:
\begin{itemize}
	\item The 6-degrees-of-freedom translation and orientation of the room in the microphone array coordinate frame;
	\item The 3-dimensional source position in the microphone array coordinate frame;
	\item The 3 dimensions of the room;
	\item Absorption coefficients for the 6 room surfaces.
\end{itemize}
We provide an open-source algorithm and extensive experimental results that suggest that the answer to this question is \textit{yes}, under a broad range of randomized input parameters, for sufficiently large microphone arrays and sufficiently high frequencies of sampling. In particular, near exact inversion is achieved, with geometrical errors in the order of millimeters and hundredth of degrees across hundreds of randomly generated rooms of size 2$\times$2$\times$2 to 10$\times$10$\times$5 meters, using a 32-element spherical microphone array of diameter 8.4~cm and a frequency of sampling of 16~kHz. These errors keep steadily decreasing when increasing the array size and sampling rate. Errors below 3~mm are also obtained in 95\% of our test cases for an 8-element non-spherical array, outperforming by an order of magnitude the well-known baseline of Dokmanic et al. \cite{dokmanic2013acoustic} to which oracle times of arrival are provided, at a fraction of the computational cost. We finally show that the parameters estimated by the proposed inverse algorithm can be fed back to the forward model to \textit{extrapolate} RIRs at any source-array placements in the room, with signal-to-error ratios above 20~dB for large enough arrays. RIR interpolation has been recently investigated in, \textit{e.g.}, \cite{bastine2023room,sundstrom2024optimal}.

The presented algorithm builds on a recently proposed method by the authors that estimates a 3D image-source \textit{point cloud} up to a given range from a discrete multichannel RIR \cite{sprunck2022gridless}. We devise here a new three-stage procedure that recovers the 18 parameters of interest from such a point cloud. First, the 3D room orientation is estimated. Second, the true source and first order image sources are labeled based on this orientation. Third, the remaining parameters are estimated based on the locations and amplitudes of labeled image sources.


The remainder of this article is organized as follows. Section~\ref{sec:sota} recalls relevant background and offers a review of the state of the art. Section \ref{sec:model} reviews the shoebox image-source forward model and the image source localization procedure used in this study. Section~\ref{sec:algo} describes the proposed room parameter recovery algorithm. Section~\ref{sec:expe} presents extensive simulated experiments and results supporting the algorithmic reversibility claim. Finally, we provide concluding remarks and perspectives in Section \ref{sec:conclusion}.

\section{Background and State of the Art}
\label{sec:sota}

The key physical phenomenon making room geometry estimation from audio measurements possible at all is that of \textit{early acoustic reflections}. When sound propagates from a source inside of a room, it is reflected on surfaces before reaching microphones. This materializes into delayed and filtered copies of the emitted signal inside the measured time-domain signals, that are commonly referred to as \textit{echoes}. The \textit{time of arrival} (TOA) of an echo at a microphone is proportional to the length of the corresponding reflected propagation path, while the \textit{time differences of arrival} (TDOAs) of an echo between two or more microphones are linked to the \textit{direction of arrival} (DOA) of the corresponding reflected propagation path. The core idea of nearly all existing methods in the field is to estimate such quantities from measured signals, to prune, sort and label echoes, and to solve for the acoustic-scene geometry based on the recovered information. A litterature review of the works tackling some or all of these steps is proposed in the remainder of this section.

The reflector associated to the TOA of a first-order propagation path from a source to a microphone is known to be tangential to an ellipsoid whose focci are the corresponding source and microphone positions. Assuming the latter are known, a number of early approaches, referred to as \textit{direct localization} in \cite{remaggi2016acoustic}, have hence focused on detecting, pruning, clustering and localizing tangent lines to multiple ellipses in the 2D case \cite{antonacci2010geometric,filos2010two,canclini2011direction,canclini2011exact,filos2012localization,antonacci2012inference}, or tangent planes to multiple ellipsoids in the 3D case \cite{nastasia2011localization,remaggi20153d,remaggi2016acoustic, macwilliam2023simultaneous}. An alternative to this is to combine the TOAs and DOAs of echoes to obtain the 3D locations of their associated image sources. Reflectors can then be localized as the bi-secting planes between a true source and its first order image sources, as in \cite{dokmanic2011can,ribeiro2011geometrically,dokmanic2013acoustic,mabande2013room,rajapaksha2016geometrical,remaggi2016acoustic,remaggi2018acoustic,lovedee2019three,tuna20203d}. This approach is referred to as \textit{image source reversion} in \cite{remaggi2016acoustic} and is the one employed in this article.

Several early works in the field assume that TOAs are trivial to estimate from room impulse responses (RIRs) using peak picking \cite{antonacci2010geometric,antonacci2012inference} or consider them readily available \cite{dokmanic2011can, nastasia2011localization, dokmanic2013acoustic,moore2013room,markovic2013estimation,jager2016room,rajapaksha2016geometrical, macwilliam2023simultaneous}. This would be the case if microphones, sources and reflectors had perfectly flat responses up to very high frequencies, but this is never true in practice. This band-limitedness results in a significant \textit{smearing} of echoes, blurring the location of their peaks and making them overlap and interfere with each other in the time domain. An analogous phenomenon occurs in the 1D and 2D DOA domains, and is reinforced by the limited diameter of microphone arrays. Interference is all the more present since echoes are, by definition, strongly correlated with each other and with direct-path signals. Due to this, the tasks of TOA, TDOA and DOA estimation of early acoustic reflections has been the focus of significant research effort. The vast majority of existing techniques proceed by some form of peak-picking over a \textit{discretized} time domain \cite{filos2010two,tervo2010estimation,canclini2011exact,filos2012localization,remaggi20153d,remaggi2016acoustic,el20173d,mabande2013room,kowalczyk2013blind,crocco2016estimation}, DOA domain \cite{canclini2011direction,sun2012localization,remaggi20153d,lovedee2019three,shlomo2021blindloc}, joint TOA-DOA domain \cite{torres2013room,remaggi2018acoustic,tuna20203d}, 3D space \cite{ribeiro2011geometrically} or ray space \cite{kuster2004acoustic,markovic2013soundfield}. To improve the separation and sharpness of objects inside such discrete grids, some methods leverage sparsity-based techniques \cite{ribeiro2011geometrically,kowalczyk2013blind,crocco2016estimation,bertin2016joint,tuna20203d,shlomo2021blindloc} or ad-hoc image processing tools \cite{torres2013room,markovic2013soundfield,remaggi2018acoustic,lovedee2019three}. Despite these efforts, operating over discrete time or space suffers from intrinsic limitations. First, the separability of peaks is fundamentally limited. This has led many authors to impose additional constraints on the geometrical setup to ensure separation.
Second, for 3D image-source localization, the required discrete-grid size grows cubically in the desired range and precision. This fundamentally limits the achievable resolution under reasonable computational constraints. For instance, in \cite{ribeiro2011geometrically}, a sparse problem over a 3D grid of 900k points needs to be solved to achieve an angular resolution of $\approx4^{\circ}$ and a distance resolution of $\approx2$~cm, while restricting the array-reflector distances to at most 3.5~m. Third, sparse optimization over a discrete grid fundamentally suffers from the so-called \textit{basis-mismatch} problem \cite{denoyelle2019sliding,chi2011sensitivity}, generally requiring the use of ad-hoc post-processing steps.

There are a few notable exceptions to this discrete grid-search paradigm \cite{markovic2013estimation,zamaninezhad2014localization,tervo2015direction,di2020blaster,sprunck2022gridless}. In \cite{markovic2013estimation}, a class of 2D room geometries is selected (rectangle, L-shaped) and the continuous shape dimensions are directly optimized by minimizing a distance between measured and image-source TOAs, using a genetic algorithm. In \cite{zamaninezhad2014localization}, the wall- and source-to-wall distances in a 1D room are continuously optimized based on resonant frequencies. In \cite{tervo2015direction}, non-linear minimization of a likelihood-based cost function in the spherical harmonics domain is utilized to jointly estimate the continuous DOA of a fixed number of reflectors. In \cite{di2020blaster}, the TDOAs of echoes are blindly estimated in the continuous time domain by leveraging an infinite-dimensional convex relaxation of the problem and the \textit{sliding Frank-Wolfe} algorithm \cite{denoyelle2019sliding}. In \cite{sprunck2022gridless}, a similar approach is employed in 3D space to recover the continuous 3D positions of all image sources within a given range from a multichannel RIR. The present work builds on this last approach.

Many of the above-reviewed methods estimate TOAs and/or TDOAs independently across individual channels and/or channel pairs. To leverage these quantities for geometry estimation, they need to be associated to reflectors, a procedure referred to as \textit{echo sorting}. This difficult combinatorial problem is the focus of \cite{dokmanic2011can,dokmanic2013acoustic,jager2016room}. The need for echo sorting is bypassed by methods that directly localize image sources from RIRs \cite{ribeiro2011geometrically,mabande2013room,tervo2015direction,remaggi2016acoustic,remaggi2018acoustic,lovedee2019three,tuna20203d,shlomo2021blindloc,sprunck2022gridless}, such as the one employed in this work.

Once image sources are localized, a necessary subsequent step is to \textit{label} them, namely, identify their order of reflection. In the literature, labeling is typically performed by ad-hoc algorithms that exploit the geometrical constraints at hands, \textit{e.g.}, \cite{ribeiro2011geometrically,dokmanic2013acoustic,rajapaksha2016geometrical,lovedee2019three, macwilliam2023simultaneous}. They are often tailored to the specific class of source-microphone-room setup under consideration, and may hence be hard to generalize. In this work, a new approach to labeling is presented. We leverage the fact that the recently proposed image-source localization method in \cite{sprunck2022gridless} can recover a much larger number of image sources than previously possible. We present a new technique that robustly estimate the 3D \textit{orientation} of the room in the microphone array frame from such image-source point cloud. This specific task has not been investigated before, to the best of the authors' knowledge. Once the room axes have been estimated, identifying first-order image sources becomes relatively straightforward, namely, they are the closest ones to the true source along each of the 6 oriented room axes.

Complementarily to approaches tackling room geometry estimation, a few approaches are focused on estimating surface absorption coefficients or echo amplitudes from recorded signals given the room geometry or image source positions \cite{nava2009situ,antonello2015evaluation,bertin2016joint,shlomo2021blind, okawa2021estimation,dilungana2022geometry}. The approaches in \cite{nava2009situ,antonello2015evaluation,bertin2016joint,okawa2021estimation} proceed by discretizing the wave equation in both time and space to solve the corresponding sparse inverse problem. The computational burden of discretizing limits these approaches to either frequencies below 500~Hz \cite{nava2009situ,antonello2015evaluation} or 2D rooms \cite{bertin2016joint,okawa2021estimation}. In contrast, the approach in \cite{shlomo2021blind} estimate echo amplitudes blindly given their continuous TOAs via least-square optimization. To tackle the high sensitivity of these techniques to geometrical errors, \cite{dilungana2022geometry} formulates the problem in the magnitude short-time Fourier domain and robustly solves the corresponding non-linear inverse problem with the help of random sampling consensus.

Finally, a relatively recent class of methods replaces some or all of the previously described steps by making use of virtually-supervised deep learning \cite{yu2020room,tuna2023data,yeon2023rgi,bicer2024data}. While promising results have been reported, these approaches are currently restricted to the acoustic setups simulated in their training data. Moreover, the ability of various training simulation strategies to generalize to a broad enough range of real measurements is an open question that calls for further investigation.

We close this section by observing that most of the above-referenced methods are only tested on a restricted set of geometries. For instance, the methods in \cite{kuster2004acoustic,antonacci2010geometric,filos2010two,tervo2010estimation,nastasia2011localization,dokmanic2011can,canclini2011direction,canclini2011exact,filos2012localization,sun2012localization,markovic2013estimation,markovic2013soundfield,torres2013room,mabande2013room,dokmanic2013acoustic,remaggi20153d,shlomo2021blindloc} are tested on less than 3 simulated or real room geometries, and the experimental setups in \cite{ribeiro2011geometrically,filos2012localization,sun2012localization,mabande2013room,remaggi2016acoustic,el20173d,remaggi2018acoustic,lovedee2019three} have in common a favorable positioning of devices, such as a microphone array near the room center, or sources near the reflectors of interest. Combined with the general absence of publicly available code, this makes existing techniques difficult to reproduce or compare. With the hope of offering a strong baseline, we open-source here an algorithm that is applicable to RIRs simulated by the image source method under fully randomized input parameters, without requiring any hyper-parameter tuning.

\section{Image Source Model and Localization}
\label{sec:model}
\subsection{The Image Source Method}
The ISM \cite{allen1979image} relies on a heuristic to approximate the wave equation with impedance boundary conditions. The partial differential equation with boundary conditions is replaced by the following free-field equation containing an image-source \textit{point cloud} as a source term:
\begin{equation}
\label{eq:IS_wave}
\frac{1}{c^2}\frac{\partial^2 p}{\partial t^2} (\rvect,t) - \Delta p(\rvect,t) = \sum_{k=0}^{+\infty} a_{k} \delta_{\rvect_k}\left(\rvect\right)\delta_0(t).
\end{equation}
Each $\rvect_k$ corresponds to the location of a point source emitting an impulse at $t=0$, casting an outward spherical wave. Each of these image-sources is equivalent to a reflection path of the original sound wave on the walls\footnote{In the remainder of the manuscript, for convenience, the term \textit{wall} is used to refer to any of the 6 room boundaries, including the floor and the ceiling}. First order image-sources are constructed by taking the symmetry of the source with respect to the walls, and higher order reflections are obtained by iterating this process. 

In the shoebox case, the image-source coordinates are most easily expressed in a \textit{reference frame} of the room, meaning a frame composed of an orthonormal basis $(\boldsymbol e_1,\boldsymbol e_2,\boldsymbol e_3)$ of normal vectors to the walls along with an origin located at one of the room's corners. 
In such a frame, the image-source coordinates are given by:
\begin{equation}\label{eq:IS_coord}
    \{\rvect_{\boldsymbol q,\boldsymbol \varepsilon} =\boldsymbol \varepsilon \odot \boldsymbol v_{d^\text{src}} + 2 \boldsymbol q\odot \boldsymbol v_L \; | \;  \boldsymbol \varepsilon \in \{-1;1\}^3,\; \boldsymbol q \in \Z^3\}
\end{equation}
where $\boldsymbol v_L=[L_x,L_y,L_z]^\top$ is the room size vector and $\boldsymbol v_{d^\text{src}}= [d_x^\text{src}, d_y^\text{src}, d_z^\text{src}]^\top$ gives the distance of the source to each wall containing the origin. Hence, the image sources lie on eight distinct translated orthogonal lattices of common mesh size $2L_x\times 2L_y\times 2 L_z$.

Each image source is weighted with an amplitude $a_k\in [0,1]$ that models the multiplicative decay caused by the reflections of the original sound wave on the walls. The amplitudes of first-order sources correspond to the reflection coefficients of each wall. 
Note that, as proven in an appendix of the original paper of Allen and Berkley \cite{allen1979image},  equation \eqref{eq:IS_wave} is only equivalent to the original wave equation with boundary conditions if each reflection coefficient is equal to one, which corresponds to the case of perfectly reflecting (rigid) boundaries, \emph{i.e}, homogeneous Neumann boundary conditions.

The analytical solution to equation \eqref{eq:IS_wave} is given by a linear combination of delayed Green functions:
\begin{equation}\label{eq:solution_wave}
    p(\rvect,t) = \sum_{k=0}^{+\infty} a_{k}\frac{ \delta\left(t-\left\|\rvect-\rvect_k\right\|_2/c\right)}{4\pi\left\|\rvect-\rvect_k\right\|_2}.
\end{equation}
In practice, $p$ is only observed at some microphone locations $\{\boldsymbol r_m^\text{mic}\}_{m=1}^M$. Moreover, the observed signal is filtered by the microphones and sampled in time, yielding a discrete measurement vector $\boldsymbol x\in \R^{MN}$, defined componentwise by:
\begin{align}
     x_{m,n} \coloneqq&   \left(\kappa_m*p(\boldsymbol r_m^\text{mic}, \cdot)\right)(n/f_s) \\
     =& \sum_{k=0}^{K} a_{k}\frac{ \kappa_m\left(n/f_s-\left\|\boldsymbol r_m^\text{mic}-\rvect_k\right\|_2/c\right)}{4\pi\left\|\boldsymbol r_m^\text{mic}-\rvect_k\right\|_2}\label{eq:forward_model}
\end{align}
where $f_s$ is the sampling frequency and $\{\kappa_m\}_{m=1}^M$ are source-microphone response filters. Most ISM implementations use ideal low-pass filters $\kappa_m(t)=\kappa(t)=\operatorname{sinc}(\pi f_s t)$, which will also be the case throughout this article.
Each sub-vector $\boldsymbol x_m\in\mathbb{R}^N$ corresponds to a discrete and filtered RIR presenting filtered spikes at the times of arrival of image sources. Note that we only consider a finite number of image sources in equation \eqref{eq:forward_model}. This approximation is reasonable because in practice synthesized RIRs are finite in time and the effect of reflections arriving at microphones later than the final time $N/f_s$ is negligible.   
%
%
\subsection{Source recovery}
\label{sec:spl}
In order to estimate image source locations from $\boldsymbol x$ we apply the algorithm previously proposed in \cite{sprunck2022gridless}, which is briefly reviewed below. The discrete multichannel RIR $\boldsymbol x$ can be expressed as the forward pass of the right hand side of equation \eqref{eq:IS_wave} through a linear observation operator $\Gamma$, \emph{i.e.}
\begin{equation}
    \boldsymbol x=\Gamma(\sum_{k=0}^{K} a_{k} \delta_{\rvect_k}).
\end{equation}
Let $\M_*(\R^3)\subset \M(\R^3)$ be the subset of \textit{Radon measures}\footnote{$\M(\R^3)$ is the topological dual of the space of continuous functions on $\mathbb{R}^3$ that vanish at infinity, see \cite{denoyelle2019sliding}.} that can be written as a linear combination of Dirac masses.
The following optimization problem is considered to jointly recover the image source locations and amplitudes:
\begin{equation}\label{eq:ls_opt_pb}
\operatorname*{argmin}_{\psi\in\M_*(\R^3)}\left\|\boldsymbol x - \Gamma\psi\right\|^2_2.
\end{equation}
This problem is non-convex, but can be relaxed to an infinite-dimensional convex problem by extending it to the whole set of Radon measure $\M(\R^3)$ and by regularizing with a total variation norm on $\psi$. This relaxed problem can then be addressed by the \emph{sliding Frank-Wolfe} algorithm, a greedy approach proposed by \cite{denoyelle2019sliding}, that belongs to the broader class of so-called \textit{super-resolution} or \textit{gridless} techniques. We adapted this method to tackle problem \eqref{eq:ls_opt_pb} in \cite{sprunck2022gridless}, and were able to accurately recover hundreds of image sources within range, for large enough sampling frequencies and array sizes. Importantly, the recovered image sources are \textit{unlabeled}, some may be missing, and false positives or mislocated sources may exist. Recovering the room parameters from such a noisy, unlabelled image source point cloud is the main contribution of this article, as presented in the following section.



\section{Recovery of Room Parameters}
\label{sec:algo}
This section presents the proposed room parameter estimation algorithm, consisting in three steps that are detailed in each of the following subsections. The first step is to estimate the room's orientation, the second step is to label the original and first-order image sources, and the third step consists in inferring the remaining parameters, \emph{i.e.}, the source position, the distance of the source relative to each wall (room translation), the room dimensions, and the wall absorption coefficients.
\blue{Overall, the assumptions made by the approach are the following:
\begin{itemize}
    \item The room is cuboid (shoebox) and empty;
    \item Reflections at the room boundaries are specular, with one unknown absorption coefficient per surface;
    \item The internal geometry of the microphone array is non-coplanar and known;
    \item The source and the microphones are omnidirectional with perfectly known responses.
\end{itemize}
}

\subsection{Room Orientation}
\begin{figure}[!t]
\vspace{-12mm}
\centering
\subfloat[]{\includegraphics[width=0.5\linewidth]{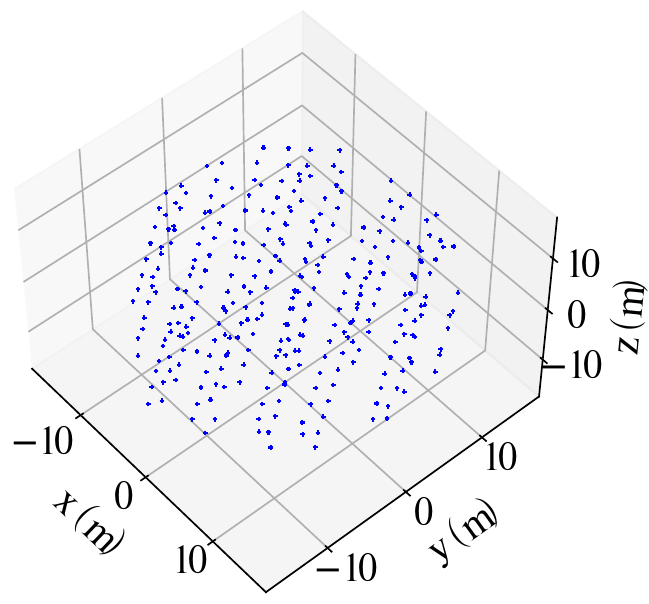}%
}
\hfill
\subfloat[]{\includegraphics[width=0.5\linewidth]{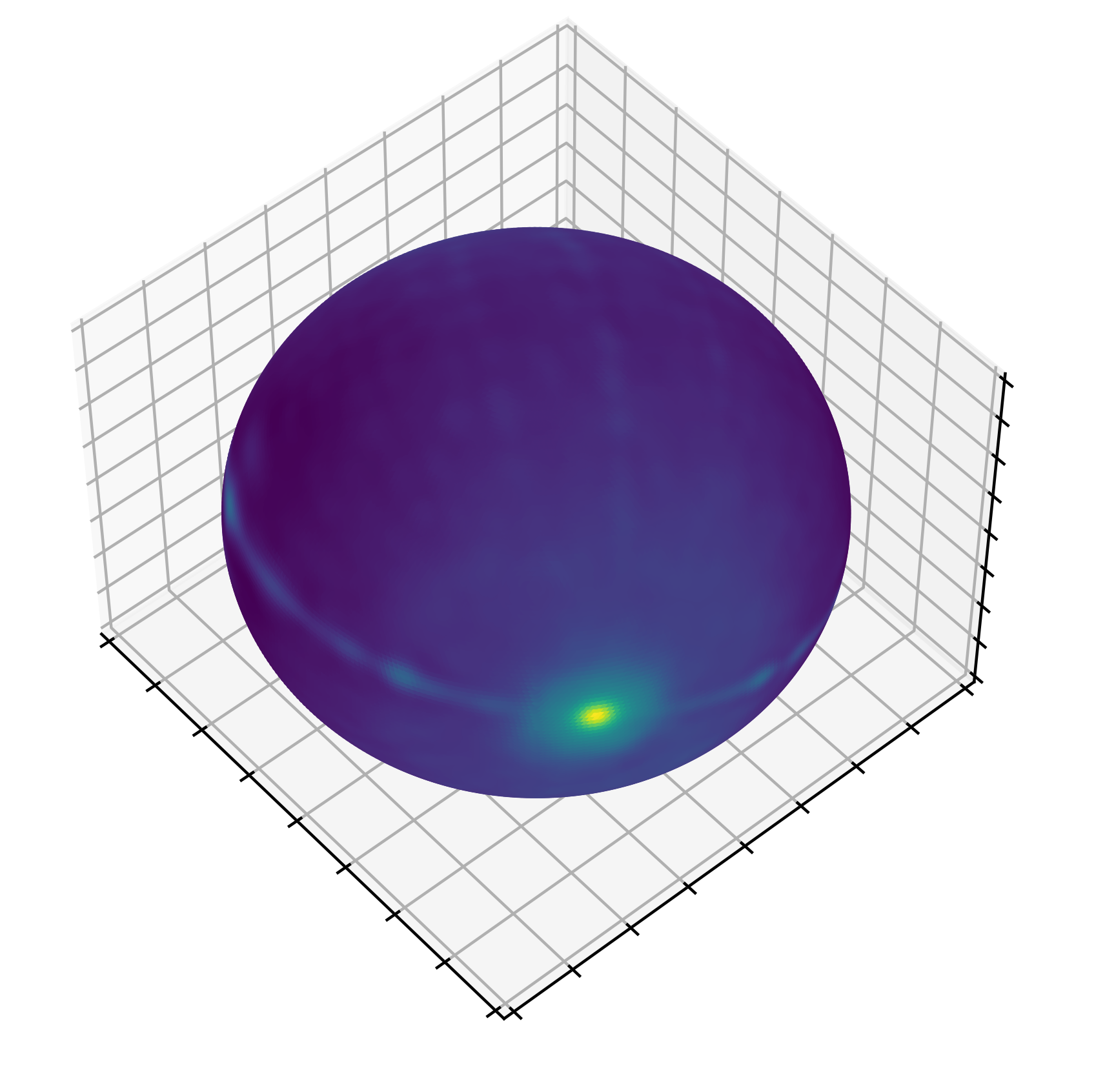}%
}
\caption{(a) Reconstructed image-source point cloud using \cite{sprunck2022gridless} (b) Associated $J_3^\sigma$ score plotted on the sphere (brighter is higher). A sharp peak is observed in the direction of a wall normal.}
\label{fig:costfun3d}
\end{figure}

\begin{figure}\vspace{-2mm}
    \centering
    \includegraphics[width=2. in]{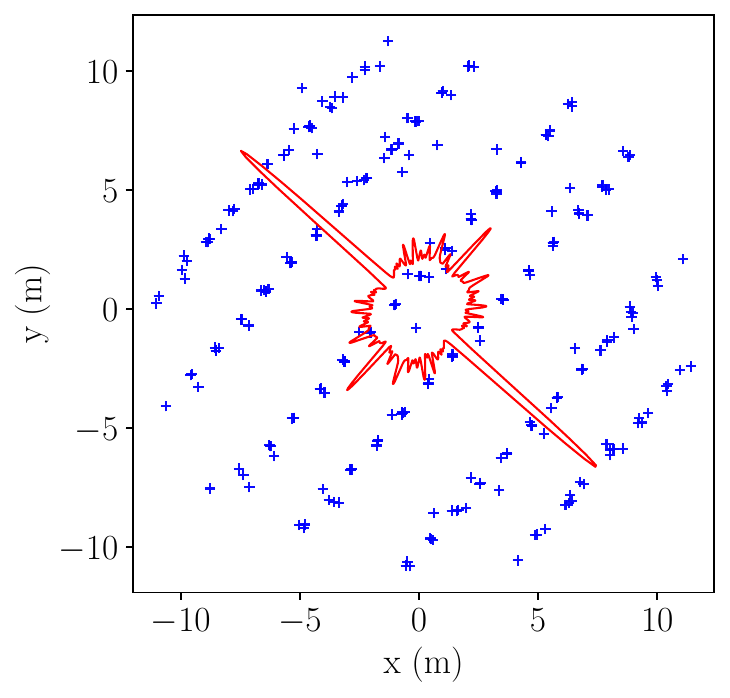}
    \caption{Projection of the estimated sources on $ \hat {\boldsymbol e}_1$ (blue) and the associated 2D $J_{2, \hat {\boldsymbol e}_1}^\sigma$ score (red).  We observe maximal values in the directions of the wall normals.}
    \label{fig:costfun2d}
    \vspace{-2mm}
\end{figure}

Let us first consider the task of recovering the room orientation from an unlabelled image source point cloud. The key idea is to estimate its underlying orthogonal grid structure, which is apparent in the examples of Fig.\ref{fig:costfun3d}\textbf{(a)} and \ref{fig:costfun2d}. The task amounts to finding a rotation matrix that transforms the microphone array's reference frame to the room's reference frame, up to a permutation of directions.
By Eq.~\eqref{eq:IS_coord}, the projected coordinates of image sources onto a normal vector to a wall will form clusters, each cluster containing the coordinates of a plane of image sources parallel to this wall. Conversely, projecting image sources onto a randomly chosen vector will, intuitively, not form clusters but instead spread out over the entire range of possible values. 
In other words, the room basis vectors are orthogonal to the image-source planes generated by the corresponding walls and are expected to maximize the number of orthogonalities.
Our method seeks to exploit this structure by scoring basis vector candidates according to their orthogonality to the directions generated by image-source pairs. 
Formally, let us define $f_D$ as follows:
\begin{equation}\label{def_f}
   \forall \boldsymbol u,\boldsymbol v \in \R^D,\quad f_D(\boldsymbol u,\boldsymbol v)=\left\{\begin{array}{cc}
                        1 &\text{ if } \boldsymbol u \bot \boldsymbol v\\
                        0 & \text{otherwise}.
                    \end{array}\right.
\end{equation}
Let $\mathcal G\subset \R^3$ be a finite set of image source locations. Let us consider the following optimization problem: 
\begin{equation}\label{opt_orientation}
\max_{\Ltn{\boldsymbol u} = 1 }J_3(\boldsymbol u), \quad \textrm{where} \quad J_3(\boldsymbol u)= \sum_{\boldsymbol s,\boldsymbol p\in \mathcal G}f_3(\boldsymbol u,\boldsymbol s-\boldsymbol p).
\end{equation}
It can be shown that in the noiseless case, for a complete finite cuboid grid $\mathcal G$ of image sources, the solution to this problem is indeed a wall normal:
\begin{proposition}\label{prop:optim_solution}
    Let $N_1, N_2, N_3$ be non-zero even integers. Consider the following subset of image sources:
    $\mathcal G=\{\rvect_{\boldsymbol q,\boldsymbol \varepsilon}, \; \boldsymbol q\in\dint{0}{N_1/2-1}\times \dint{0}{N_2/2-1} \times \dint{0}{N_3/2-1}  ,\; \varepsilon_i \in \{-1,1\}\}$ with $\rvect_{\boldsymbol q,\boldsymbol \varepsilon}$ defined as in \eqref{eq:IS_coord}. Then, \blue{every} solution $\boldsymbol u^*$ to problem (\ref{opt_orientation}) is a wall normal, \emph{i.e}, $\boldsymbol u^*=\pm \boldsymbol e_i$ for some $i\in\dint{1}{3}$. 

    \noindent Proof:  See Appendix \ref{sec:app}.
\end{proposition}
%
Note that in Proposition \ref{prop:optim_solution}, the coordinates are expressed using Eq.~$\eqref{eq:IS_coord}$, \textit{i.e.}, in the unknown reference frame of the room. However, the definition of the cost function $J_3$ is independent of the coordinate system, so that the result remains true in any coordinate frame. Note also that adversarial cases could be built by carefully removing sources from the image-source point cloud in order to have the score function bear its maximum in a wrong direction. 
However, assuming the reconstruction algorithm of Section \ref{sec:spl} misses image sources at random, the probability of encountering such an adversarial situation is vanishingly small, and Proposition \ref{prop:optim_solution} is expected to hold for generic subsets, as will be confirmed by our experiments. 
\begin{algorithm}[t]  
\caption{Orientation estimation}\label{alg:alg1}
\begin{algorithmic}[1]
\REQUIRE{Image sources $(\rvect_k)_{k=1}^K$}
\ENSURE{Estimated room orthonormal basis $\hat{\boldsymbol e}_1,\hat{\boldsymbol e}_2,\hat{\boldsymbol e}_3$}
\STATE $\hat{\boldsymbol e}_1 \gets \operatorname{argmin}_{\boldsymbol u \in \mathcal{S}^2_\text{discr}}J_3^{0.01}$
\FOR{$\sigma \in [0.01, 0.005, 0.0005]$}
   \STATE $\hat{\boldsymbol e}_1 \gets \operatorname{local\_descent}(\hat{\boldsymbol e}_1, J_3^\sigma)$
\ENDFOR
\STATE $\hat{\boldsymbol e}_2 \gets \operatorname{argmin}_{\boldsymbol u \in \mathcal{S}^1_\text{discr}}J_{2,\hat{\boldsymbol e}_1}^{0.01}$
\FOR{$\sigma \in [0.01, 0.005, 0.0005]$}
   \STATE $\hat{\boldsymbol e}_2 \gets \operatorname{local\_descent}(\hat{\boldsymbol e}_2, J_{2,\hat{\boldsymbol e}_1}^\sigma)$
\ENDFOR
\STATE $\hat{\boldsymbol e}_3 = \hat{\boldsymbol e}_1 \times \hat{\boldsymbol e}_2$
\end{algorithmic}
\end{algorithm}

In practice, the image-source reconstruction is noisy while the function $f_D$ defined in \eqref{def_f} \blue{only captures \textit{exact} orthogonalities}. $f_D$ is instead approximated using a Gaussian kernel
\begin{equation}
   f^\sigma_D(\boldsymbol u,\boldsymbol v)=\exp\left(-\frac{1}{2\sigma^2}\left(\frac{\boldsymbol u.\boldsymbol v}{\Ltn{\boldsymbol u}\Ltn{\boldsymbol v}}\right)^2\right),
\end{equation}
such that $\lim_{\sigma\rightarrow 0}f^\sigma_D = f_D$ in the pointwise sense. The scale parameter $\sigma$ controls the tightness of the approximation and 
plays a regularizing role with respect to the error committed in the localization of image sources. A small $\sigma$ will yield a noisy loss function if the source localization error is high. Conversely, a large $\sigma$ means poor precision on room orientation recovery.
As we are searching for an optimal \emph{unit} vector, the regularized score function $J^\sigma_3$ can be re-parameterized in spherical coordinates by two angles $(\theta,\phi)\in[0,2\pi[\times[0,\pi[$:
\begin{equation}
    J^\sigma_3(\theta,\phi)=\sum_{\boldsymbol s,\boldsymbol p\in \mathcal G}f^\sigma_3(\boldsymbol u(\theta,\phi),\boldsymbol s-\boldsymbol p)
\end{equation}
where $\boldsymbol u(\theta,\phi)$ is the unit vector defined by spherical coordinates $(\theta,\phi)$. Once a first basis vector $\boldsymbol u$ maximizing $J_3^\sigma$ has been found, we can proceed in a greedy manner by projecting $\mathcal G$ onto $\boldsymbol u^\bot$:
\begin{equation}
    J^\sigma_{2,\boldsymbol u}(\theta)=\sum_{\boldsymbol s,\boldsymbol p\in \mathcal G}f^\sigma_2(\boldsymbol v(\theta),\mathcal P_{\boldsymbol u^\bot} (\boldsymbol s-\boldsymbol p))\quad \forall \theta\in[0,2\pi[.
\end{equation}
As can be seen in the examples of Fig.~\ref{fig:costfun3d}\textbf{(b)} and \ref{fig:costfun2d}, both score functions $J^\sigma_3$ and $J^\sigma_{2,\boldsymbol u}$ feature maxima along the room axes.
We use the \emph{Scipy} implementation of the \emph{BFGS} algorithm \cite{2020SciPy-NMeth} to maximize  $J^\sigma_3$. Due to the non-convexity of the problem we initialize the optimization algorithm on a finely meshed half-sphere $\mathcal{S}^2_\text{discr}$. In order to reduce even more the chance of the algorithm stopping at a local minimum, we begin with a high value of the scale parameter $\sigma$ and perform the optimization with gradually decreasing values. This process yields an accurate, gridless reconstruction of a first basis vector $\hat {\boldsymbol e}_1$, given a sufficiently accurate image-source reconstruction. The sources are then projected onto the plane orthogonal to $\hat {\boldsymbol e}_1$ and the process is repeated to recover a second vector $\hat {\boldsymbol e}_2$ by optimizing $J^\sigma_{2,\hat {\boldsymbol e}_1}$. The third vector is then obtained by taking the cross product $\hat {\boldsymbol e}_1\times \hat {\boldsymbol e}_2$. The full process is summarized in Algorithm \ref{alg:alg1}. The same values of $\sigma$ will be used in all of the experiments in this article, without any specific tuning.

\subsection{First order identification and geometry inference}
\begin{algorithm}[t]
\caption{Source-wall distances, first order amplitudes}\label{alg:alg2}
\begin{algorithmic}[1]
\REQUIRE{Image sources and amplitudes $(\boldsymbol r_k)_{k=1}^K, (a_k)_{k=1}^K$; directions $\hat {\boldsymbol e}_1,\hat {\boldsymbol e}_2, \hat {\boldsymbol e}_3$; threshold $\mu$}
\ENSURE{Corrected amplitudes up to order 1 $\hat a_0,\ldots, \hat a_6$, source-walls distances $\hat d_1,\ldots,\hat d_6$}
\STATE $\hat {\boldsymbol r}_0 \gets \operatorname{fusion}(\boldsymbol r_{k_0}, (a_k)_k, (\boldsymbol r_k)_k, \mu), \; k_0 = \operatorname{argmin}_{k}\Ltn{\boldsymbol r_k}$
\FOR{$t=1,\ldots,3$}
\STATE $(\boldsymbol r_\text{left},\boldsymbol r_\text{right})\gets \operatorname{closest\_in\_cone}(\hat {\boldsymbol r}_0, \hat {\boldsymbol e}_t, (\boldsymbol r_k)_k)$
\STATE $(\hat a_{t-}, \hat {\boldsymbol r}_{t-}) \gets \operatorname{fusion}(\boldsymbol r_\text{left}, (a_k)_k, (\boldsymbol r_k)_k, \mu)$
\STATE $(\hat a_{t+}, \hat {\boldsymbol r}_{t+}) \gets \operatorname{fusion}(\boldsymbol r_\text{right}, (a_k)_k, (\boldsymbol r_k)_k, \mu)$
\ENDFOR
\end{algorithmic}
\end{algorithm}

Once the room orientation has been estimated, we seek to identify which of the estimated image sources are of first order. We leverage the fact that the zero-th order image source, \emph{i.e.}, the true source, can be straightforwardly identified. Indeed, it is necessarily the closest one to the microphone array's center. It is also accurately localized, since the direct path is generally well separated from reflections in RIRs.
We then cast a cone from the true source in each reconstructed direction $\hat {\boldsymbol e}_d$ and their opposite  $-\hat {\boldsymbol e}_d$. The image source closest to the true source within each cone is picked as a first-order candidate. If the cone is empty (implying that source localization errors are too great) we progressively extend the cone's width until it contains at least one source. As the reconstruction algorithm sometimes produces clusters of sources around the true image-source locations, we assume that any source close to an estimated first order source is a reconstruction artifact. We thus proceed to merge the closest estimated sources. Let $\boldsymbol r^*$ be a candidate first-order source, $\mu \in \R_+^*$ a threshold, and $\{\boldsymbol r^*_1,\ldots \boldsymbol r^*_P \}$ the set of reconstructed sources such that $\Ltn{\boldsymbol r^*_p-\boldsymbol r^*} < \mu \; \forall p\in \dint{1}{P}$. We use a heuristic inspired by \cite{traonmilin2020projected} to merge the corresponding Diracs and their amplitudes:
\begin{equation}
    \hat{a} = \sum_{p=1}^P a_p^*,\quad
    \hat{\boldsymbol r} = \sum_{p=1}^P \frac{a_p^*}{\hat{a}}\boldsymbol r_p^* .
\end{equation}
This procedure gives us estimates for the six first-order image sources and their associated reflection coefficients.
The distances of the true source to each wall are then recovered by computing the projections on each estimated wall normal. Let $\hat \rvect_{t-}, \hat \rvect_{t +}$ be the first order image sources corresponding to $\hat {\boldsymbol e}_t$ such that $\hat \rvect_{t+}$ is in the cone emitted from $\hat \rvect_0$ with direction $\hat {\boldsymbol e}_t$ and $\hat \rvect_{t-}$ is contained in the opposite cone. The room length in that direction is given by the following formula:
\begin{equation}
  \hat{L}_t =
  \hat {\boldsymbol e}_t.(\hat {\boldsymbol r}_{t+}-\hat {\boldsymbol r}_{t-})/2.
\end{equation}
Setting the intersection of the walls corresponding to $\hat {\boldsymbol r}_{1-}, \hat {\boldsymbol r}_{2-}, \hat {\boldsymbol r}_{3-}$ as a reference vertex of the room, the translation vector of the room with respect to the source is:
\begin{equation}
    \hat{\boldsymbol \tau}_{\text{room}}= \frac{1}{2}\begin{pmatrix} \hat {\boldsymbol e}_1 .(\hat {\boldsymbol r}_0 - \hat {\boldsymbol r}_{1-}) \\ \hat {\boldsymbol e}_2 .(\hat {\boldsymbol r}_0 - \hat {\boldsymbol r}_{2-}) \\ \hat {\boldsymbol e}_3 .(\hat {\boldsymbol r}_0 - \hat {\boldsymbol r}_{3-})\end{pmatrix}.
\end{equation}
Given the coordinates $\rvect$ of a point in the frame of the microphones, we can then compute the corresponding coordinates in the recovered room frame:
\begin{equation} \label{eq:inv_formula}
    \hat \rvect_\text{room} = 
    \begin{pmatrix}\hat {\boldsymbol e}_1^T \\ \hat {\boldsymbol{e}}_2^T \\\hat {\boldsymbol e}_3^T\end{pmatrix} (\rvect-\hat \rvect_0) +  \hat{\boldsymbol \tau}_{\text{room}}.
\end{equation}
%


We now have recovered all 18 input parameters that were used to generate the multichannel RIR:
\begin{itemize}
\item the room orientation $\hat{ \boldsymbol{e}}_1,\hat {\boldsymbol{e}}_2,\hat {\boldsymbol{e}}_3$,
\item the 3D source position $\hat{\rvect}_0$,
\item the room translation with respect to the source $\hat{\boldsymbol \tau}_\text{room}$,
\item the room dimensions $\hat L_1,\hat L_2, \hat L_3$,
\item the 6 wall absorption coefficients $\hat{\alpha}_k = 1- \hat{a}_k^2$ for $k=1,\dots,6$.
\end{itemize}
Our open-source code for the full image-source reversion procedure is available at: \urlstyle{tt}\url{https://github.com/Sprunckt/acoustic-sfw}.

\section{Numerical Experiments}
\label{sec:expe}
We proceed in this section to evaluate the effectiveness of the proposed  inverse algorithm, which can be decomposed into two major steps: first estimating an image source point cloud from a multi-channel RIR and then inferring the room parameters from it. The first step was extensively tested in \cite{sprunck2022gridless}, so we focus here on the estimation of the 18 room parameters given an image-source point cloud estimated using \cite{sprunck2022gridless}, as described in Sec.~\ref{sec:spl}. All the following tests are based on RIRs simulated using the \blue{vanilla} shoebox ISM. \blue{We used our own implementation}\footnote{\blue{Our implementation uses true $\operatorname{sinc}$ functions while \emph{pyroomacoustics} uses approximations to save computational time.}} \blue{of Eq. \eqref{eq:forward_model} with the image-source locations generated by the \emph{pyroomacoustics} package \cite{scheibler2018pyroomacoustics}.} As in \cite{sprunck2022gridless}, \blue{image sources up to reflection  order $20$ are simulated and the RIRs are cut} after $50~$ms, so that all audible reflections are present in the signals. 

\subsection{Simulation Details}

\begin{figure}
    \centering
    \includegraphics[width=1\linewidth]{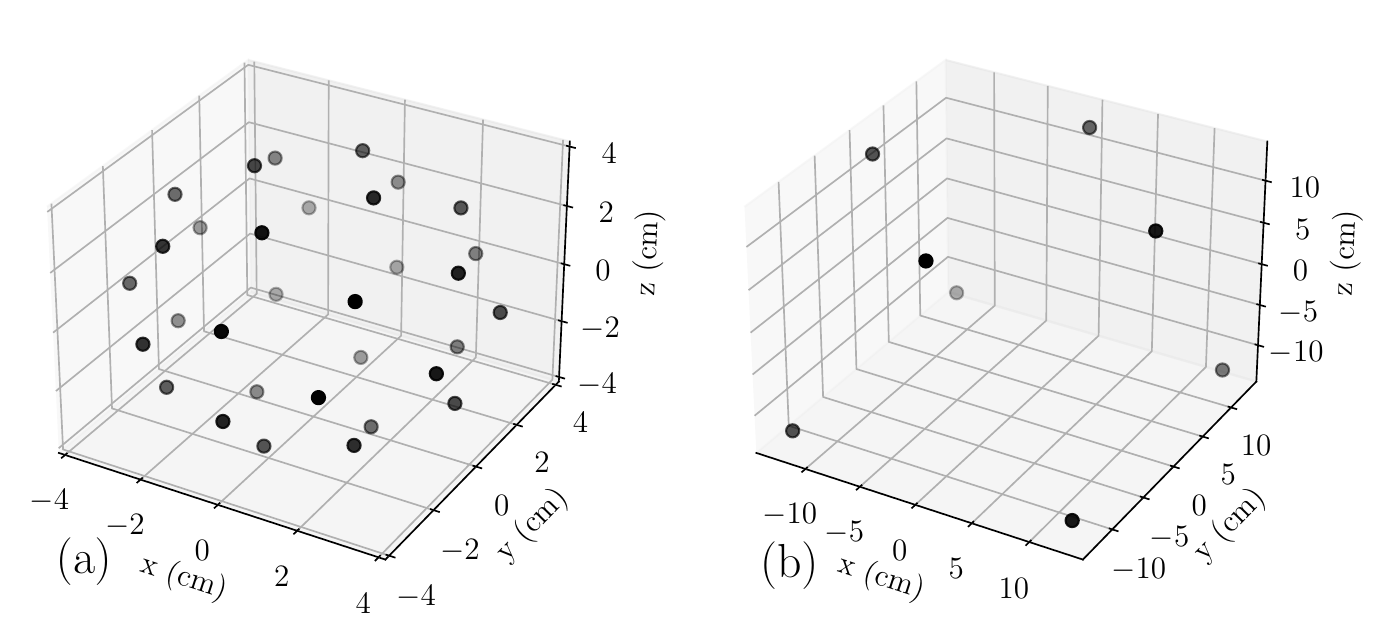}
    \caption{\blue{Geometries of the microphone arrays used in the experiments. (a) Smallest spherical array used in Section~\ref{subsec:results}, (b) non-spherical array used for the baseline comparison in Section~\ref{subsec:comparison}}}
    \label{fig:mic_pos}
\end{figure}
\label{sec:sim}
We test the full reconstruction procedure on a set of $200$ randomly generated rooms containing an omnidirectional impulse sound source and a microphone array. We use two distinct array geometries that are \blue{depicted in Fig.~\ref{fig:mic_pos} and} detailed in the following subsections. 
The rooms' lengths and widths in meters are picked uniformly at random in $[2,10]$ while the heights are picked in $[2,5]$\blue{, encompassing most commonly encountered shoebox room sizes, from tiny water closets to large halls and long corridors}. The array's center and the source are randomly placed in each room and the array is randomly rotated. \blue{A minimal separation distance of 1~m between the source and the array is enforced, as per common measurement practice, to avoid having the direct path completely dominate the RIR}.  We also enforce a distance constraint of $25$~cm of the array center to each wall to avoid having any microphone placed beyond the room's boundary. Each wall absorption coefficient is drawn uniformly at random in $[0.01, 0.3]$, \blue{which is representative of most commonly encountered surface materials in rooms (see, \textit{e.g.}, Fig.~1 in \cite{foy2021mean}). While specifically treated acoustic panels may feature higher absorption coefficients in practice, we purposefully exclude those to make the result analysis clearer. Indeed, highly absorbent walls could simply become \textit{inaudible} in RIRs, making their localization trivially ill-posed, and calling for a dedicated separate analysis.}

\subsection{Evaluation Metrics}
\label{sec:metrics}
\subsubsection{Orientation and dimensions}
In order to match each recovered direction with the corresponding ground truth wall normal, we apply the ground truth inverse rotation to $(\hat {\boldsymbol{e}}_1,\hat {\boldsymbol{e}}_2, \hat {\boldsymbol{e}}_3)$. Each resulting vector should contain two zero coefficients, the last coefficient being $-1$ or $1$. The indices of the non-zero coefficients allow us to re-order the vectors of the rotation matrix to match the recovered directions. We then compute the mean angular errors between the recovered directions $\hat e_d$ and the associated wall normals by taking the arccosine of the dot products. We also compute the mean absolute errors on recovered room dimensions.
\subsubsection{Wall absorptions}
Having matched recovered first-order sources to walls, we compute the mean absolute errors on estimated absorption coefficients $\hat{\alpha}_{1:6}$.
\subsubsection{Room translation}
In order to evaluate the room translation estimation, we calculate the room's center in the array's reference frame from estimated parameters.
%
This is done by inserting $\hat {\boldsymbol r}_\text{room}=[\hat L_1/2, \hat L_2/2, \hat L_3/2]^\top$ in \eqref{eq:inv_formula} and solving for $\rvect$.
We then calculate the mean of Euclidean distances to the ground truth. 

\subsubsection{RIR extrapolation}
Lastly, we evaluate the global accuracy of the method by re-simulating a RIR $\hat{\boldsymbol x}$ corresponding to a new random source-array placement in the room using the image-source method \eqref{eq:forward_model} with estimated parameters as input. 
Using the same sampling rate, we compute the signal-to-error ratio to the true RIR $\boldsymbol x$ at the new location:
\begin{equation}
\operatorname{SER}(\hat{\boldsymbol x},\boldsymbol x)=10\log_{10}\left(\frac{\sum_{i=1}^{NM}x_i^2}{\sum_{i=1}^{NM}(\hat x_i- x_i)^2}\right).
\end{equation}
\subsection{Experimental Results and Analysis}
\label{subsec:results}

\begin{figure}[t]
    \centering
    \includegraphics[width=\linewidth]{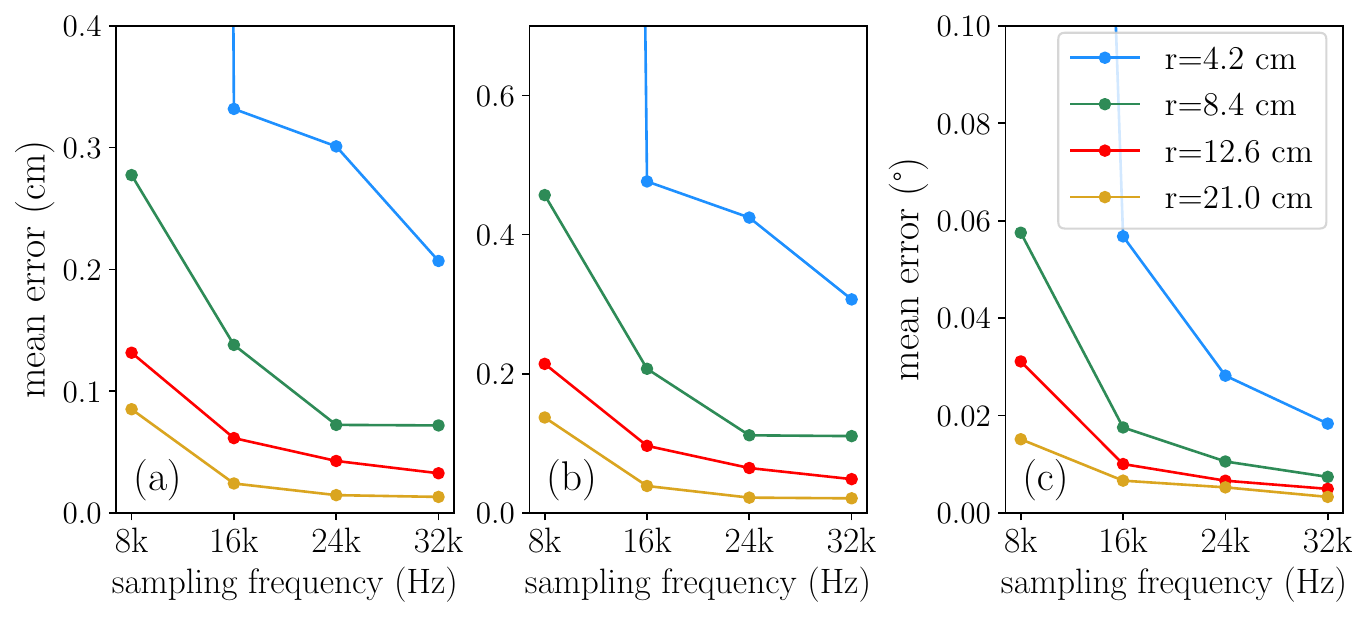}
    \caption{Mean absolute errors on room dimensions (a), mean Euclidean errors on room center (b) and mean angular error on room orientation (c) in function of the sampling frequency for varying array radii and frequency of sampling.}
    \label{fig:mean_dim}
\end{figure}
We first consider a $32$-element spherical microphone array based on the geometry of the em32 Eigenmike\textsuperscript{\textregistered} (radius $R=4.2$~cm) and scaled by various factors.
Figure \ref{fig:mean_dim} presents the algorithm's performance on the geometry estimation task for varying sampling frequencies and microphone array radii. In accordance with the image source localization results reported in \cite{sprunck2022gridless}, the accuracy of the estimation improves as the radius or the sampling frequency grow. The lowest resolution ($R=4.2$~cm and $f_s=8$~kHz) \blue{yields a few reconstruction failures} that heavily impact the mean errors. \blue{For instance, $3.5~\%$ of the errors on wall distance estimation are over $50$ cm in that case.} 
These \blue{large-error cases vanish for all the larger array sizes and sampling rates considered}, the mean error steadily converging towards zero for all three metrics. This empirically supports our main claim that the shoebox image-source method is indeed fully algorithmically reversible for large enough arrays and frequencies of sampling.
For a frequency of sampling of $24$~kHz and the lowest radius, the mean room dimension estimation error is \blue{already very low}, around $3$~mm. This number goes down to $0.15$~mm when dilating the array by a factor of 5.
Meanwhile, the mean error on room orientation (Figure \ref{fig:mean_dim}.c) remains under $0.06^{\circ}$ in all experiments, except for the very lowest resolution.
The errors on room center localization \blue{are slightly higher, at $4.2$~mm with the smallest array at $24$~kHz and $0.22$~mm after $5\times$ dilation}. This increase is expected because estimating the room center couples errors on orientation estimation and source-wall distance estimation.

\begin{figure}[t]
    \centering
    \includegraphics[width=\linewidth]{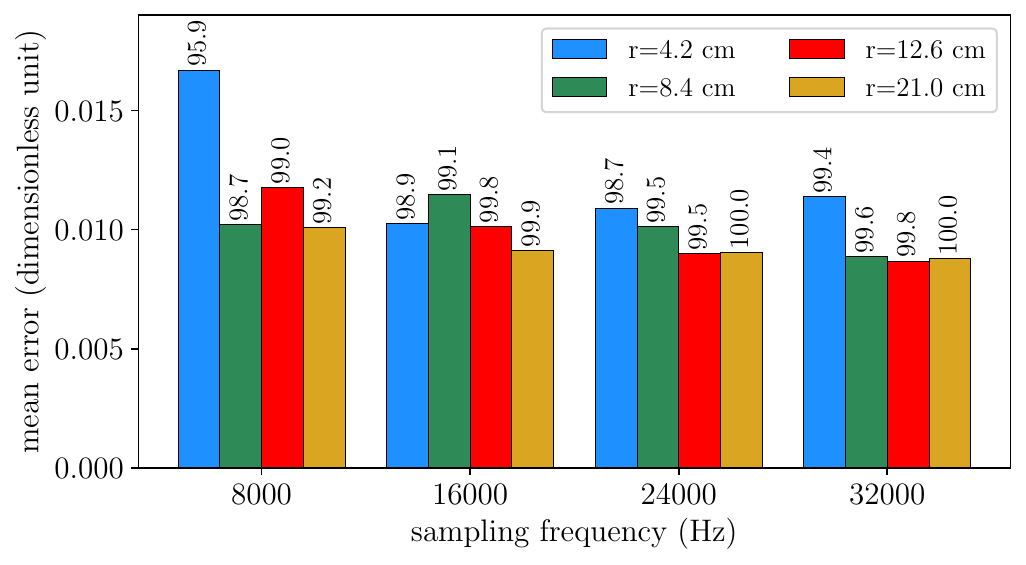}
    \caption{Mean absolute error on absorption coefficients recovered below a $0.3$ threshold for varying array radii and frequency of sampling.  The recall for this threshold is indicated above each bar, in percent.}
    \label{fig:mean_abs}
            \vspace{-3mm}
\end{figure}

We then evaluate the estimation of wall absorption coefficients. \blue{Here, we observed some rare ($<1\%$) failures of absorption recovery even for relatively high array resolutions and frequency of sampling.}
In order to get a more meaningful picture of the error committed, we thus only compute the mean errors over coefficients estimated with an error below $0.3$, \blue{and consider the rest as outliers (recall that in our simulations, the coefficients take values in $[0.01, 0.3]$).}
We also compute the recall rates for this threshold. Both metrics are displayed in Fig. \ref{fig:mean_abs}. 
The obtained mean errors are around $0.01$ with a $100\%$ recall rates for the largest array and frequencies of sampling above 24~kHz. While these are very low errors, we do not observe the same convergence towards zero as on geometrical errors. One possible explanation is that we kept the spike estimation algorithm described in \cite{sprunck2022gridless} untouched, including two spike pruning steps that discard low amplitude Diracs before and after the final gradient descent. While the first pruning step does seem to help the optimization algorithm, the second step, which aimed at reducing false positives, might cause an issue on amplitude estimation. Rather than deleting the spikes and losing the corresponding amplitudes, a lead for improvement would be to merge the spikes by, \emph{e.g.}, adapting the heuristic presented in \cite{traonmilin2020projected}.

\begin{figure}[t]
    \centering 
    \includegraphics[width=\linewidth]{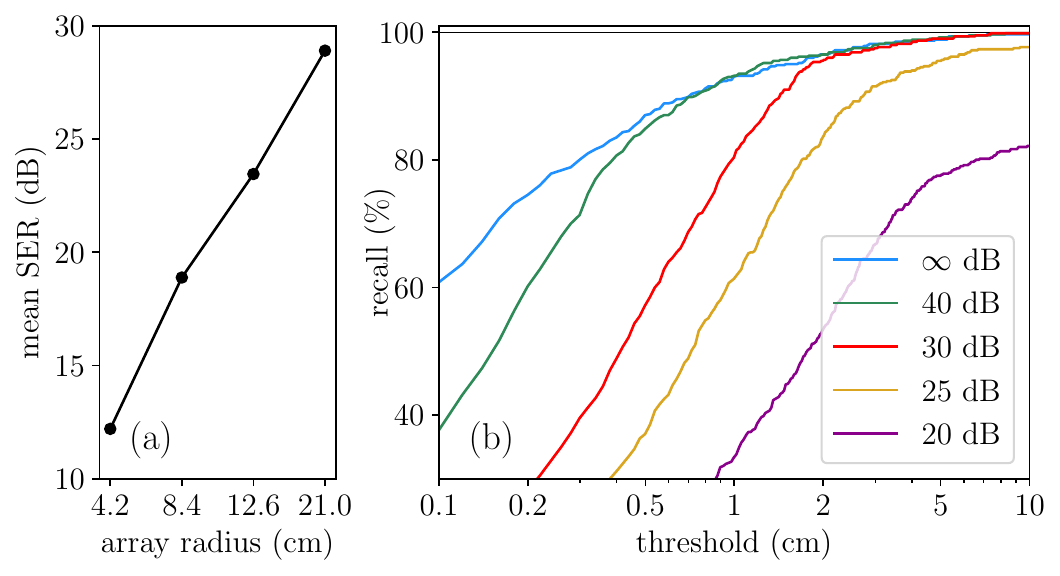}
    \caption{(a) Mean signal-to-error-ratio of RIR extrapolation for varying array radii at $f_s=24$~kHz (b) Recall on room dimension recovery as a function of threshold for varying PSNRs for an array radius $r=4.2$~cm and a frequency of sampling $f_s=24$~kHz.}
    \label{fig:rir_conv}
\end{figure}

\begin{figure}[t]
    \centering
    \includegraphics[width=\linewidth]{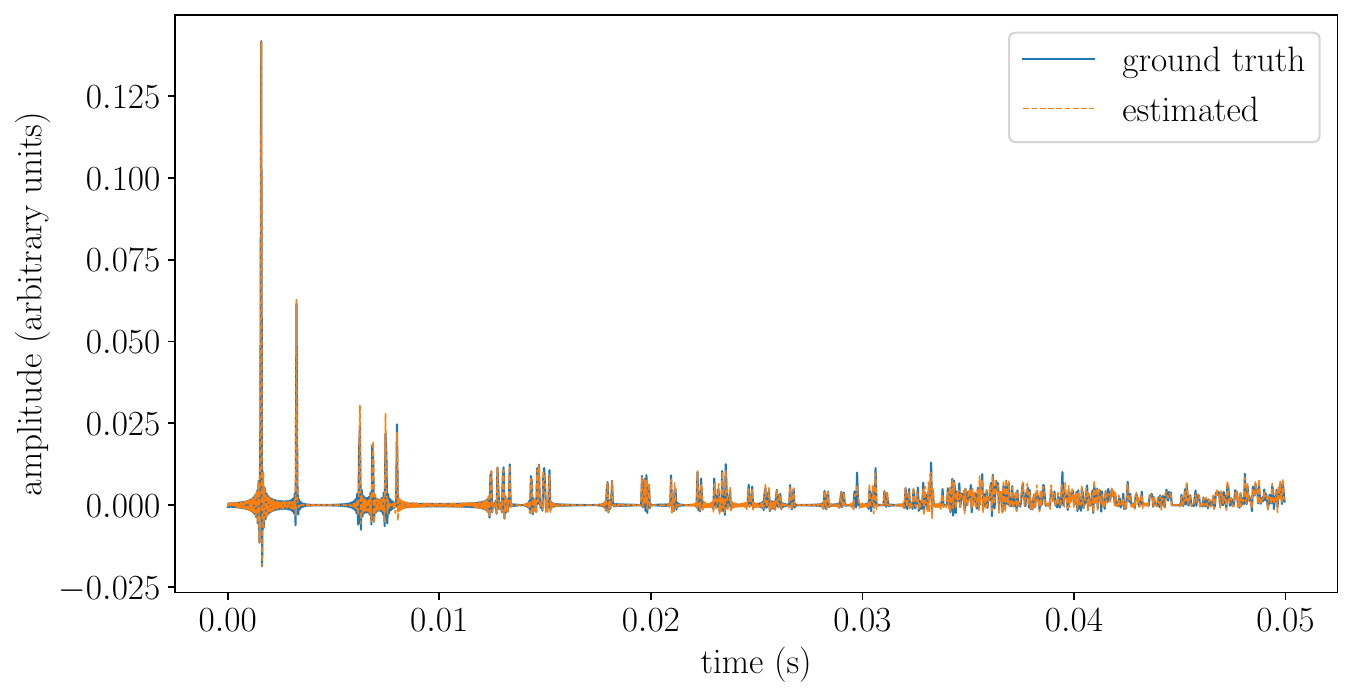}
    \caption{Example of RIR extrapolation inside the room of Fig. \ref{fig:costfun2d} ($4.2$~cm array radius, $24$~kHz frequency of sampling).}
     \label{fig:rir_reconstruction}
     \vspace{-2mm}
\end{figure}

We now proceed with evaluating the ability of the method to extrapolate RIRs to arbitrary source-array placements in the same room. The results are shown in Fig.~\ref{fig:rir_conv}(a).
We again observe a strong convergence of RIR extrapolation errors towards zero as the array size increases, bringing further support to the claim that the shoebox image-source method has been successfully reversed. 
Note that we did not observe such convergence as a function of the frequency of sampling. This is expected, since the RIR extrapolation task itself, as assessed by the proposed metric, becomes harder as the frequency of sampling increases. 
An example of RIR extrapolation result is presented in Fig.~\ref{fig:rir_reconstruction}. As can be seen, the extrapolated RIR very closely matches the ground truth.

We finally study the impact of noise on room size estimation. The sampling frequency and array radius are respectively set to $24$~kHz and $4.2$~cm, and we proceed to varying the peak signal-to-noise ratio (PSNR) of input signals using additive white Gaussian noise \blue{uncorrelated across channels}. \blue{This is meant to coarsely emulate generic signal degradation but is not representative of typical degradations present in measured RIRs. Note also that the use of PSNR occludes the fact that in a RIR, the peaks of first-order echoes, all of which are necessary for room-size estimation, are typically much lower than the global peak. To give an idea of this, in each RIR from the 25-dB-PSNR test set, the PSNR of the weakest first order echo taken in isolation ranges from 1.5 dB to 19~dB (11.7~dB on average), which constitutes a significant degradation from a signal processing perspective.} As expected, the algorithm's performance deteriorates when the noise increases and a severe drop appears \blue{below $20$~dB PSNR}. Nevertheless, the algorithm still manages to recover $95.5\%$ of all room dimensions with an error below $5$~cm \blue{under a PSNR of $25$~dB}, suggesting a reasonable robustness of the approach.

\subsection{Baseline Comparison}
\label{subsec:comparison}
We now compare the accuracy of the proposed algorithm with the landmark Euclidean distance matrix (EDM)-based method introduced by Dokmanic et al. \cite{dokmanic2013acoustic}, using the code provided by the authors\footnote{\urlstyle{tt}\url{https://infoscience.epfl.ch/record/186657/}}. This method takes as input a set of unlabelled times of arrival (TOAs) on multiple RIRs, and returns the 3D locations of first order image sources. Direct comparison on the synthetic dataset defined in Section \ref{sec:sim} turned out to be unfeasible.  Indeed, the \blue{computational cost (in time and memory)} of the EDM-based method explodes when the number of reflections is too large. Moreover, the method makes the strong assumption that only TOAs from image sources of orders lower than or equal to two are provided. Even when only considering these low-order sources, the number of considered combinations can become very high if the reflections are tightly clustered together due to the room's configuration, which frequently happens in our dataset. \blue{Finally, we observed that the computational cost of the method also drastically increases with the number of channels. In particular, applying the algorithm to the $32$-channel spherical array used in other experiments, or even to arrays of more than 8 microphones, was not feasible without running out of time and memory (in \cite{dokmanic2013acoustic}, arrays of 5 microphones were used). To reach a fair compromise between the computational feasibility and the performance of this baseline}, we consider a non-spherical microphone array of 8 microphones \blue{consisting of} two squares stacked on top of each other, the top square being rotated by an angle of $\pi/4$. The corresponding array diameter is $37.5~$cm \blue{and the array geometry is depicted in Fig.~\ref{fig:mic_pos}}(b). \blue{This configuration is also meant to illustrate the applicability of our approach to a different array geometry and number of microphones}. In order to avoid choosing a peak-picking technique to process the input of the EDM-based method, we place it in an oracle setting. Namely, we provide it with the true times of arrival of all image sources up to order 2 that are in recording range (partial oracle labeling), rounded to the nearest discrete-time sample at 32 kHz.
Note that working in discrete time is a fundamental limit of such approaches. We run the two algorithms on the same room configurations as before, only altering the array's geometry but retaining the same location for its center.

\begin{table}[b]
    \centering
    \caption{Recall, precision and mean Euclidean errors (MEE) for \blue{correctly recovered} first-order image sources ($\text{O}_1$) and MEE for the true source ($\text{O}_0$) using \cite{dokmanic2013acoustic} or the proposed method.}
    \label{tab:baseline_comparison}
    \begin{tabular}{c|c|c|c|c}
         &  $\text{O}_1$  Rec. & $\text{O}_1$  Prec. & $\text{O}_1$  MEE & $\text{O}_0$ MEE\\
        \hline
        \cite{dokmanic2013acoustic} & 84.4$\%$ & 59.7$\%$ & 65.7$\pm$ 41.3 mm & 35.1$\pm$ 26.0 mm\\
        Ours & 97.2$\%$ & 97.2$\%$ & 2.41$\pm$ 5.71 mm& 0.289$\pm$ 0.584 mm 
    \end{tabular}
\end{table}

For each method, we compute the precision and recall for a $20$~cm error threshold on the source and first-order image sources localization and labelling. While the proposed algorithm always returns exactly $6$ first-order sources, the EDM-based method can wrongfully label second-order reflections as first-order reflections, causing a loss in precision. The results for these experiments are listed in Table \ref{tab:baseline_comparison}. The localization errors obtained with the EDM-based method are \blue{over} an order of magnitude larger than with the proposed approach. This highlights that, even using oracle information, the considered task is far from trivial when considering fully randomized room parameters. The proposed algorithm obtains a mean Euclidean error below 3~mm, which is below $\frac{343}{2\times 32000}\approx 5.1~$mm, the theoretically lowest achievable radial error by any discrete-time method at this frequency of sampling, indicating that super-resolution is achieved. The number of rooms for which all $6$ first-order sources were retrieved without spurious second-order ones was $25.5 \%$ for the EDM-based method.
Hence, the method could not be used to recover the full geometry of most of the rooms.
In contrast, this ratio reached $95.5 \%$ of the rooms using the proposed method.
%
%
For those rooms, the mean geometrical reconstruction errors obtained by it, following the metrics presented in Section \ref{sec:metrics}, were respectively $0.34\pm 0.6$~mm for the room dimensions, $0.61\pm 0.6$~mm for the room translation and $0.016\pm 0.05^\circ$~mm for the room orientation.
These are in line with those obtained with the $32$-element spherical microphone array of comparable radius and sampling frequency. This seems to indicate that when the array resolution is sufficient, adding microphones does not significantly improve the accuracy of correctly recovered sources. However, adding microphones does seem to reduce some of the geometrical ambiguities and hence to increase the number of correctly identified sources.

\section{Conclusion and Perspective}
\label{sec:conclusion}
A new algorithm that leverages the gridless image-source localization method introduced in \cite{sprunck2022gridless} to achieve full image-source reversion from a discrete, low-passed, multichannel, shoebox RIR was presented. Extensive numerical experiments on simulated RIRs from randomized input parameters reveal that near-exact recovery of all input parameters is achieved by the method, for large enough array sizes and sampling rates. This constitutes, to our knowledge, the first empirical evidence that the historical image-source method of Allen and Berkley \cite{allen1979image} is algorithmically reversible, for a wide range of configurations.

The proposed approach is currently not directly applicable to real measured RIRs. This is mainly because the image source localization method it relies upon is specifically designed to reverse the forward image-source model, which makes a number of simplifying assumptions that do not hold in reality. A path towards real-data applicability can nevertheless be envisioned. For this, the method in \cite{sprunck2022gridless} would need to be extended to take into account both angular and frequency dependencies of receiver, source, and wall responses. Even assuming the responses of the source and microphones are known, and using a physics-based model for the angular dependencies of wall responses, the number of unknown in the problem is then significantly increased. Namely, one needs to additionally estimate the source (and image sources) orientations, as well as a frequency-dependent impedance for each wall. Leveraging additional geometrical and physical constraints on these unknown or incorporating stochastic models are promising leads to make the corresponding inverse problem tractable. Another avenue for future research is to go beyond shoebox geometry. This requires tackling the problem of occlusions, namely, that some image sources may not be visible by all microphones. It also makes the task of room orientation recovery more difficult, and calls for the development of more general point-cloud-to-geometry techniques.

{\appendices
\section{Proof of Proposition \ref{prop:optim_solution}}

\begin{proof}
\label{sec:app}
Note that, by construction, $|\mathcal G|=N_1N_2N_3$.  
Let $\boldsymbol u\in\R^3$ and denote by $\mathcal P^{\boldsymbol u}_{\boldsymbol s}$ the affine plane passing by $\boldsymbol s\in\mathcal G$ with normal vector $\boldsymbol u$. $J_3$ can be reinterpreted as the total number of intersections of all planes $\{\mathcal P^{\boldsymbol u}_{\boldsymbol s}\}_{\boldsymbol s\in\mathcal G}$ with $\mathcal G$:
     \begin{equation}
        J_3(\boldsymbol u) = \sum_{\boldsymbol s\in\mathcal G}\vert\mathcal G \cap \mathcal P^{\boldsymbol u}_{\boldsymbol s}\vert.
    \end{equation}
 Indeed, for all $\boldsymbol s,\boldsymbol p\in\mathcal G$, $\boldsymbol s-\boldsymbol p$ is orthogonal to $\boldsymbol u$ if and only if $\boldsymbol p\in \mathcal P^{\boldsymbol u}_{\boldsymbol s}$. 
 Note that for $1\leq i\leq 3$ the set $\mathcal G$ is partitioned by the disjoint union of $N_i$ parallel planes $\mathcal Q_j^i\coloneqq \mathcal P^{\boldsymbol e_i}_{\boldsymbol s^i_j},\; 1\leq j\leq N_i$ where $\{ \boldsymbol{s}_j^i, 1\leq j \leq N_i\}=\{\rvect_{\boldsymbol q, \boldsymbol \varepsilon}\in\mathcal G, (q_l,\varepsilon_l)=(0,1) \text{ if } l\neq i\}$.
    Then:
    \begin{equation}\label{eq:partition_L}
        J_3(\boldsymbol u) = \sum_{\boldsymbol s\in\mathcal G}\sum_{j=1}^{N_i}\vert\mathcal Q_j^i \cap \mathcal P^{\boldsymbol u}_{\boldsymbol s}\cap \mathcal G\vert\quad \forall i \in  \dint{1}{3}.
    \end{equation}
   
Assume in the following that $\boldsymbol u$ is not colinear to any of the vectors $\boldsymbol e_i$. Then there exists a direction $\boldsymbol e_i$ such that every line $\mathcal Q^i_j \cap \mathcal P_{\boldsymbol s}^{\boldsymbol u}$, $1\leq j\leq N_i$ is diagonal, in the sense that the direction of the line is not given by any of the basis vectors $\boldsymbol e_j$. 
Indeed, consider the converse proposition by contradiction: assume that for each $i \in \dint{1}{3}$ there exists an image source $\boldsymbol s\in\mathcal G$ and a plane $\mathcal Q_j^i$ such that the line $\mathcal Q_j^i \cap \mathcal P^{\boldsymbol u}_{\boldsymbol s}$ is generated by a basis vector $\boldsymbol e_{k_i}$, $k_i\neq i$. In particular, $\boldsymbol u$ is orthogonal to $\boldsymbol e_{k_1}$ by definition as $\mathcal P^{\boldsymbol u}_{\boldsymbol s}$ contains the direction $\boldsymbol e_{k_1}$. Similarly, $\boldsymbol e_{k_{k_1}}$ is orthogonal to $\boldsymbol u$. Moreover, as the direction $\boldsymbol e_{k_{k_1}}$ is contained in $\mathcal{Q}^{k_1}_j$ which is orthogonal to $\boldsymbol e_{k_1}$, then $\boldsymbol e_{k_1}$ and $\boldsymbol e_{k_{k_1}}$ are distinct. Hence $\boldsymbol u$ would be colinear to the last basis vector, raising a contradiction. 
    
We can assume without any loss of generality that direction $\boldsymbol e_1$ verifies this property (see Fig. \ref{fig:intersection} for a depiction in that case), \emph{i.e} every line $\mathcal Q^1_j \cap \mathcal P_{\boldsymbol s}^{\boldsymbol u}$, $1\leq j\leq N_1$ is not generated by $\boldsymbol e_2$ or $\boldsymbol e_3$.
Then, the line $\mathcal Q_j^1  \cap \mathcal P^{\boldsymbol u}_{\boldsymbol s}$ intersects $\mathcal G$ at at most $\min (N_2,N_3)$ image sources. Moreover, as $\boldsymbol u$ is not colinear to $\boldsymbol e_2$ and $\boldsymbol e_3$, this upper bound can only be reached for the sources $\boldsymbol s$ located on the diagonal. Indeed, we need only consider the worst-case scenario, in which the straight line passes through all the nodes on the diagonal. These nodes are at most $\min(N_2,N_3)$.
Hence:
\begin{figure}[!t]
    \begin{center}
        \includegraphics[scale=1]{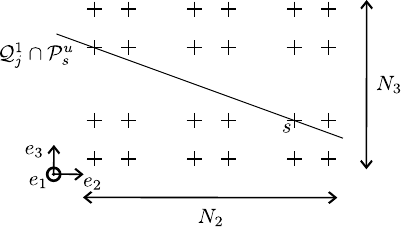}
    \end{center}
        
    \caption{Intersection of $\mathcal Q_j^1$ and $\mathcal P_{\boldsymbol s}$ when $\boldsymbol u$ and $\boldsymbol e_1$ are not orthogonal}
    \label{fig:intersection}
    \vspace{-3mm}
\end{figure}
\begin{align}
\label{eq:ineq_L}
    \!\!\!\!J_3(\boldsymbol u) \!= \!\sum_{\boldsymbol s\in\mathcal G}\!\sum_{j=1}^{N_1}\!\vert\mathcal Q_j^1 \cap \mathcal P^{\boldsymbol u}_{\boldsymbol s}\cap\mathcal G\vert &\!<\! \sum_{\boldsymbol s\in\mathcal G}\!N_1\!\min(N_2,N_3)\nonumber\\
    &\!=\!N_1^2N_2N_3\min(N_2,N_3).
\end{align}
Now by replacing $\boldsymbol u$ with $\boldsymbol e_1$, formula \eqref{eq:partition_L} becomes :
\begin{equation}
    J_3(\boldsymbol e_1) = 
    \sum_{j=1}^{N_1}\sum_{\boldsymbol s\in\mathcal G} \vert\mathcal Q_1^1\cap \mathcal G \vert\mathds{1}_{\boldsymbol s\in \mathcal Q_j^1}=
    N_1\vert\mathcal Q_1^1\cap \mathcal G \vert^2.
\end{equation} 
Thus, $J_3(\boldsymbol e_1) = N_1N_2^2N_3^2$.
We obtain similar formulas for $\boldsymbol e_2$ and $\boldsymbol e_3$, hence:
\begin{equation}
    J_3(\boldsymbol u) < \max_{1\leq i\leq 3} J_3(\boldsymbol e_i) = N_1N_2N_3 \max_{1\leq i< j\leq 3}N_iN_j
\end{equation}
and the maximum is reached for a vector $\boldsymbol{e}^*$ colinear to $\boldsymbol e_1$, $\boldsymbol e_2$ or $\boldsymbol e_3$. Note that this proof extends to 2D by considering the projection of $\mathcal G$ on $\boldsymbol{e}^{*\bot}$ in order to obtain a second basis vector.
\end{proof}

\bibliographystyle{IEEEtran}
\bibliography{biblio}

\vfill

\end{document}